\documentclass{physeauth}

\usepackage{ifthen}
\usepackage{ifpdf}

\usepackage{graphicx}
\usepackage{amsmath}
\usepackage{amssymb}
\usepackage{bm}

\ifpdf
\usepackage{epstopdf}
\else
\usepackage{epsfig}
\fi

\newcommand{\const}{\mbox{const}}

\newcommand{\eexp}{\mbox{e}^}

\newcommand{\tbox}[1]{\mbox{\tiny #1}}

\newcommand{\amatrix}[1]{\begin{matrix} #1 \end{matrix}}

\newcommand{\be}[1]{\begin{eqnarray}\ifthenelse{#1=-1}{\nonumber}{\ifthenelse{#1=0}{}{\label{e#1}}}}
\newcommand{\ee}{\end{eqnarray}} 
\newcommand{\beq}{\begin{eqnarray}}
\newcommand{\eeq}{\end{eqnarray}}

\newcommand{\hide}[1]{}

\newcommand{\putgraph}[2][width=0.30\hsize]{\includegraphics[#1]{#2}}

\begin{document} 

\begin{frontmatter}

\title{Quantum transport and counting statistics \\ in closed systems\thanksref{thank1}}

\author{Maya Chuchem, Doron Cohen}

\address{Department of Physics, Ben-Gurion University, Beer-Sheva 84105, Israel}

\thanks[thank1]{Physica E {\bf 42}, 558-563 (2010). Special issue. \\ Proceedings of FQMT conference (Prague, 2008).}

\begin{abstract}
A current can be induced in a closed device by changing control parameters. 
The amount $Q$ of particles that are transported via a path of motion, 
is characterized by its expectation value $\langle Q \rangle$, 
and by its variance $\mbox{Var}(Q)$. We show that quantum mechanics 
invalidates some common conceptions about this statistics. 
We first consider the process of a double path crossing, 
which is the prototype example for counting statistics in 
multiple path non-trivial geometry. We find out that contrary 
to the common expectation, this process does not lead to 
partition noise.  
Then we analyze a full stirring cycle that consists 
of a sequence of two Landau-Zener crossings. 
We find out that quite generally {\em counting} statistics 
and {\em occupation} statistics become unrelated, 
and that quantum interference affects them in different ways.
\end{abstract}


\end{frontmatter}


\section{Introduction}

Consider a closed isolated quantum system, say a 3~site ring 
as in Fig.1. Quite generally, in the absence of magnetic field,  
the stationary states of the system carry zero current.
If one wants to have a non-zero current~$I$ through a section of 
the device, one has either to prepare it in a non-stationary 
state or to drive the system. Driving means changing some parameters
in time. During a time period~$t$ the amount of particles  
that get through the section is $Q$. 
One may ask what is the distribution of the measured~$Q$, 
and in particular what is the expectation value $\langle Q\rangle$, 
and what is the variance $\mbox{Var}(Q)$. 
This is known as {\em counting statistics}~\cite{levitov,nazarov,blanter}.

Typically the driving is periodic, and $Q$ is defined 
as the amount of particles that are transported per 
period. The feasibility to have non-zero~$Q$ 
(non zero ``DC" current) due to periodic (``AC") driving 
is known in the context of open geometry as ``quantum pumping" \cite{BPT,Brouwer}. 
We use the term ``quantum stirring" \cite{pml,pmx} 
in order to describe the analogous effect with regard 
to a closed device~\cite{pmc}.

\begin{figure}[t]
\begin{center}
\putgraph[width=0.65\hsize]{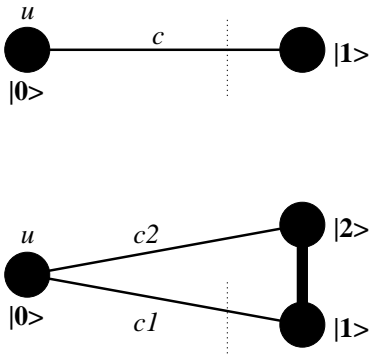}
\end{center}
\caption{ 
Toy models that are analyzed in this paper: 
a particle in a 2~site system (upper illustration), 
and a particle in a 3~site system (lower illustration).
Initially the particle is prepared in the $|0\rangle$ site
where it has a potential energy~$u$.   
The hopping amplitudes per unit time (the~$c$s) 
are also indicated. In the case of a 3~site system, 
the time units are chosen such that the hopping amplitude  
per unit time between $|1\rangle$ and $|2\rangle$ equals unity,  
while the other amplitudes are assumed 
to be small (${|c_1|,|c_2| \ll 1}$). 
The current is measured through the dotted section.} 
\end{figure}

The theory of quantum pumping and counting statistics 
in open geometries is well studied. 
Any attempt to adopt ideas from this literature 
to the present context of quantum stirring  
is dangerous, and likely to result in {\em major misconceptions},
which following~\cite{cnz,cnb} we would like to highlight. 
For this purpose let us consider the simplest model that 
can be imagined: a particle in a 3-site system (Fig.1). 
We assume that we have a control over the potential~$u$ 
of the left site, and over the couplings $c_1$ and $c_2$ 
that bind the left site to the right sites. 
The transported~$Q$ is measured through one of these two bonds. 
Let us introduce two simple questions that will be answered later on. 

{\bf Question 1. --} 
We start with a very negative~$u$ and prepare 
a particle in the left site. Then we gradually 
raise~$u$ so as to have an {\em adiabatic} transfer of 
the particle to the right side. 
If we had ${c_1 \ne 0}$ but ${c_2=0}$
one obviously expects in the strict 
adiabatic limit ${Q=1}$ with zero variance. 
We ask: what would be the corresponding result 
if we have ${|c_1|=|c_2|}$, and more generally how 
does the result depend on the relative size of the couplings? 
If the $|c_1|$ coupling is larger, does the result 
reflect having (say) ${Q=1}$ with $70\%$ probability 
and ${Q=0}$ with $30\%$ probability?

{\bf Question 2. --} 
During a cycle a conventional pump takes an electron 
from the left lead and ejects it to the right lead. 
Hence the pumped charge per cycle for a leaky pump 
is ${Q < 1}$. Now we integrate this quantum pump 
into a closed circuit and operate it. 
We ask what is the statistics of~$Q$ in the new
(integrated) configuration. Can it be much different?  
What are the maximal ${\langle Q \rangle}$ 
and minimal ${\mbox{Var}(Q)}$ that can be achieved 
per cycle? Are they both proportional to the number 
of the cycles as in the classical reasoning, 
leading to ${\sim}\sqrt{t}$ signal to noise ratio?

{\bf Outline. --} 
In the next section we define the model  
and the counting operator $\mathcal{Q}(t)$. 
In sections~3 and~4 we discuss the 
restricted quantum-classical correspondence 
that applies to the analysis of 
single path crossing, while in sections~5 
we consider multiple path geometries.  
How to treat interference in a sequence 
of Landau-Zener crossings and the analysis 
of quantum stirring are discussed in section~6.  
The long time counting statistics 
is discussed in sections~7 and~8.  
The relation to the theory of spreading and dissipation  
is illuminated in section~9. The main observations 
are summarized in section~10.

\begin{figure}[t]
\begin{center}
\putgraph[width=0.88\hsize]{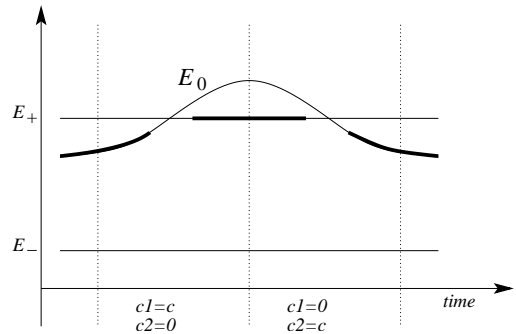}
\end{center}
\caption{
The adiabatic levels of the 3-site Hamiltonian 
during one period of a pumping cycle. 
In the absence of coupling (${c_1=c_2=0}$) 
the ${E_0=u(t)}$ level intersects  
the symmetric ${E_{+}=1}$ level. 
With non zero coupling these intersections 
become avoided crossings, and the 
particle follows adiabatically the thickened lines. 
For presentation purpose we indicate that 
either $c_1$ or $c_2$ equal zero (``blocked"), 
but in the general analysis we allow any splitting ratio, 
including the possibility ${c_1=c_2}$ of having 
the same amplitude to take either of the two paths.} 
\end{figure}

\section{Modeling}

We consider the 3~site system of Fig.1 
which is described by the Hamiltonian 
\be{1}
\mathcal{H} = 
\left( \begin{array}{ccc} 
 {u} &  {c_1} &  {c_2} \\  {c_1} & 0 &  {1} \\  {c_2} &  {1} & 0  
\end{array}\right)
\ee
We label the sites by ${i=0,1,2}$.  
We have control over the potential~${E_{0}=u}$ 
of the left site ($\#0$). 
The two right sites ($\#1$ and $\#2$) are permanently   
coupled to each other forming a double well 
with energy levels $E_{-}=-1$ and  $E_{+}=+1$. 
We also have relatively small 
couplings (${|c_1|^2,|c_2|^2 \ll 1}$) 
that allow transitions between the 
left and the right sites.

Later we consider the stirring cycle which is 
described in Fig.2.  Its operation is inspired 
by the common peristaltic mechanism. 
Namely, the coupling constants $c_1$ and $c_2$ 
are regarded as ``valves".
In the first half of the cycle ${c_2=0}$ and~$u$ is raised,  
leading to an adiabatic transfer from-left-to-right 
via the ${0\mapsto1}$ bond.
In the second half of the cycle ${c_1=0}$ and~$u$ is lowered,
leading to an adiabatic transfer from-right-to-left 
via the ${0\mapsto2}$ bond.
The net effect is to pump one particle per cycle.

The matrix representation of the operator 
which is associated with the current through 
the ${0\mapsto1}$ bond is 
\be{2}
\mathcal{I} =\left( \begin{array}{ccc} 
0 & i {c_1} & 0 \\ -i {c_1} & 0 & 0 \\ 0 & 0 & 0  
\end{array} \right)
\ee
In the Heisenberg picture the time dependent 
current operator is defined as 
$\mathcal{I}(t)= U(t)^{\dag} \mathcal{I} U(t)$,  
where $U(t)$ is the evolution operator.  
Consequently the counting operator is defined as  
\be{0}
\mathcal{Q} \ \ = \ \ \int_0^t \mathcal{I}(t')dt'
\label{Qt}
\ee
The counting operator, unlike the current operator, 
is {\em not} a conventional observable in quantum mechanics. 
What can be measured in practice are only the first 
two moments of $\mathcal{Q}$, 
which are  $\langle\mathcal{Q}\rangle$ and ${\mbox{Var}(Q)}$.
Still, on the mathematical side, it is convenient 
to treat $\mathcal{Q}$ the same way as one treats 
conventional observables. Namely, to regards its  
eigenvalues $Q_r$ as the possible outcomes of a measurement, 
and to associate with a given preparation $\psi$  
a probability distribution 
\be{4539}
\mbox{P}(Q_r) \ \ = \ \ |\langle Q_r|\psi\rangle|^2 
\ \ \ \ \ \ \ \ \ \ \ \mbox{[naive!]}
\ee
We shall refer to Eq.(\ref{e4539}) as the {\em naive} definition of the 
full counting statistics. 
The physical definition of $\mbox{P}(Q)$ is much more 
complicated \cite{levitov,nazarov}, but it leads to the same first and 
second moments \cite{cnz}. In the present paper we are not interested 
in the {\em full} counting statistics, but only in the 
first two moments, and therefore, for presentation purpose, 
we adopt the {\em naive} point of view.

\section{Single path crossing}

Let us consider first the very simple case 
of 2-site systems with left site ($\#0$) and
right site ($\#1$). 
The particle is prepared in the left site 
and after some time has some probability~$p$ 
to be found in the right site. 
This probability can be regarded as the 
expectation value of the occupation 
operator $\mathcal{N}$ that has the eigenvalues $0$ or $1$ 
depending on whether the particle is in state $\#0$ or state $\#1$.

In the classical analysis the possible outcomes 
of a measurement are ${N=Q=1}$ if the 
particle goes from left to right, and ${N=Q=0}$ otherwise.   
Obviously one should find out that 
\be{3232}
\mbox{P}(Q) = 
\left\{ 
\begin{array}{cc} 
p \ \ \ \ \ \ \ \ & \mbox{for} \ Q=1 \\
1{-}p  \ \ \ \ \ \ \ \ & \mbox{for} \ Q=0 
\end{array}
\right.
\label{Pcl}
\ee
It follows that the $k$th moment 
is ${\langle \mathcal{Q}^k \rangle = p}$ 
and therefore
\be{0}
\langle \mathcal{Q} \rangle \ \ &=& \ \ p 
\\
\mbox{Var}(Q)  \ \ &=& \ \ (1-p)p 
\ee

In the quantum mechanical treatment the counting operator 
is the integral over the current operator which has zero trace, 
so it should also be a $2{\times}2$ traceless Hermitian matrix 
which we write in the ${i{=}0,1}$ basis as   
\be{0}
{\mathcal{Q} } \ \ = \ \ 
\left( \begin{array}{cc}  
+Q_{\parallel} & iQ_{\perp} \\
-iQ_{\perp}^* & -Q_{\parallel}
\end{array} \right)
\ee
If the particle is initially prepared in the $\#0$
site then ${\langle \mathcal{Q} \rangle = Q_{\parallel}}$ 
and ${\mbox{Var}(Q)=|Q_{\perp}|^2}$.  
So now the question arises whether 
one should expect an agreement between 
the quantum result and the classical 
results for the first and second moments.
In the next section we argue that for a 
single path geometry the answer is {\em yes}: 
The first and second moments of $Q$ 
should both equal~$p$. 
It follows that the matrix that 
represents the counting operator 
in the ${i{=}0,1}$ basis is expressible 
in terms of $p$ and an extra phase:
\be{5555}
\mathcal{Q} = 
\left( \begin{array}{cc}  
+p & i\sqrt{(1{-}p)p} \ \eexp{i\phi} \\
-i\sqrt{(1{-}p)p} \ \eexp{-i\phi} & -p
\end{array} \right)
\ee
Later we are going to use this expression as 
a building block in the analysis of 
a multiple crossing scenario.

For completeness we note that from the above 
analysis it follows that the eigenvalues 
of the counting operator 
are ${Q_{\pm}= \pm\sqrt{p}}$ with 
\be{4576}
\mbox{P}(Q) = 
\left\{ 
\begin{array}{cc} 
(1{+}\sqrt{p})/2 \ \ \ \ \ \ \ \  \mbox{for} \ Q=Q_{-} \\
(1{-}\sqrt{p})/2 \ \ \ \ \ \ \ \  \mbox{for} \ Q=Q_{+}
\end{array}
\right.
\label{Pqm}
\ee
which should be contrasted with Eq.(\ref{e3232}).
The proper analysis \cite{cnz} of the {\em full} counting statistics 
gives a more complicated quasi-distribution that  
neither agrees with the naive nor with the classical result, 
but still has the same first and second moments.

The above analysis has assumed nothing about the 
detailed form of the $2\times2$ Hamiltonian. We merely had 
assumed that the particle was prepared in site $\#0$, 
and that after time $t$ there is some probability~$p$
to find the particle in site $\#1$. 
However, in the later sections we are going to discuss 
specifically adiabatic crossings. During such a process 
the on-site energies (${E_0{=}u(t)}$ and ${E_1{=}\const}$) 
cross each other,  and due to the inter-site coupling (${=}c$) 
the particle is adiabatically transferred from $\#0$ to $\#1$. 
Still there is some small probability for a non-adiabatic 
transition, so called Landau-Zener transition \cite{zener},  
that would leave the particle in $\#0$.
There is a well known formula, that allows to calculate 
the probability of such transition: 
\be{0}
P_{LZ} \ \ = \ \ \exp{\left[ -2\pi\frac{c^2}{\dot u}  \right]}
\ee
Consequently the probability to find the particle
in the right side at the end of the process is 
\be{543}
p \ \ = \ \ 1-P_{LZ}
\ee
Note that it becomes $100\%$
in the strict adiabatic limit.

\section{Quantum-classical correspondence}

For a single path transition we can prove that the first two moments 
of $\mathcal{Q}$ should be in agreement with the classical expectation.
This is based on the relation between the occupation 
operator $\mathcal{N}$ (whose eigenvalues are $0$ and $1$), 
and the counting operator $\mathcal{Q}$ (whose eigenvalues do 
not have a classical interpretation).
The relation is implied by the Heisenberg picture equation of motion:
\be{0}
\frac{d}{dt} \mathcal{N}(t) \ \ = \ \ \mathcal{I}(t) 
\ee 
Integrating over time we get
\be{0}
\mathcal{N}(t)-\mathcal{N}(0) = \mathcal{Q}
\ee
Assuming that the particle is initially 
prepared in the left site we get  
for the $k=1,2$ moments of $\mathcal{Q}$ 
\be{1515}
\langle \mathcal{Q}^k \rangle  
= \left\langle (\mathcal{N}(t)-\mathcal{N}(0))^k \right\rangle  
= \langle \mathcal{N}^k \rangle_t = p
\ee
It is important to realize that the derivation cannot 
be extended to the ${k>2}$ moments because $\mathcal{N}(t)$ 
does not commute with $\mathcal{N}(0)$.
In fact it is not difficult to calculate 
the higher moments: one just has to realize 
that from Eq.(\ref{e5555}) it follows 
that $\mathcal{Q}^2 = p\bm{1}$,  
and consequently the even moments are~$p^{k/2}$, 
while the odd moments are~$p^{(k{+}1)/2}$.
Optionally this result can be obtained 
from Eq.(\ref{e4576}) as in \cite{cnz}.

In the single path transition problem we say that 
we have {\em restricted} quantum-classical correspondence.  
The first two moments come out the same 
as in the classical calculation. Encouraged by this 
observation let us speculate what should be the results 
in more complicated circumstances that involve 
multiple path geometries...
 
Let us consider our 3~site system (Fig.1). 
We would like to analyze the first half of 
the cycle which is described by Fig.2.
The particle is initially prepared in the left site.
We gradually raise~$u$ so as to have an {\em adiabatic} transfer 
of the particle to the right side. 
The occupation probability of the right side 
at the end of the process is denoted by~$p$.  
But now we have to remember that the particle 
could get there either via the ${0\mapsto1}$ bond 
or via the ${0\mapsto2}$ bond. 
Motivated by a stochastic point of view 
one may argue that the process is like {\em partitioning} 
of a current, and therefore  
\be{0}
\langle \mathcal{Q} \rangle \ \ &=& \ \ \lambda p
\\ 
\label{e6666}
\mbox{Var}(Q)  &=&  (1-\lambda p) \lambda p 
\ \ \ \ \ \  \ \ \ \mbox{[stochastic]}
\ee 
where the splitting ratio is 
\be{6667}
\lambda \ \ = \ \ \frac{|c_1|^2}{|c_1|^2+|c_2|^2}
\ \ \ \ \ \  \ \ \ \mbox{[stochastic]}
\ee 
If for example we have a strict adiabatic process
with ${p=1}$ and the splitting ratio is $\lambda=1/2$, 
then the stochastic expectation is to have ${\mbox{Var}(Q)=(1/2)^2}$. 
Furthermore, considering a multi-cycle stirring process, 
one may argue that the variance should 
be accumulated in a stochastic manner:    
\be{0}
\mbox{Var}(Q) \ \ \propto  \ \ \mbox{time} 
\ \ \ \ \ \  \ \ \ \mbox{[stochastic]}
\ee 
All the results above that are labeled as `stochastic' 
might apply in the case of a non-coherent processes, 
or in the case of open geometries with leads attached 
to equilibrated reservoirs. 
Below we are going to show that the above `stochastic' 
results do not apply to the analysis of coherent transport 
in multiple path closed geometries.

\section{Double path adiabatic crossing}

For ${c_1=c_2=0}$ the 3~site Hamiltonian of Eq.(\ref{e1})  
is diagonal in the $|E_0\rangle$, $|E_{-}\rangle$, $|E_{+}\rangle$ 
basis. If we limit ourselves to processes which involve 
adiabatic crossings of $|E_0\rangle$ and $|E_{+}\rangle$, 
as in Fig.2, then transitions to $|E_{-}\rangle$ can be neglected 
and we can work with a  $2\times2$ Hamiltonian 
in the $|E_0\rangle$ and $|E_{+}\rangle$ representation 
\be{0}
\mathcal{H} 
\ \ = \ \ \left( \begin{array}{cc}  {u(t)} &   {c}
\\  {c} &  {1}  \end{array} \right)
\ee 
where $c=(c_1{+}c_2)/\sqrt{2}$. In the same 
representation the current operator of Eq.(\ref{e2}) 
takes the form
\be{0}
\mathcal{I}
\ \ = \ \ {\lambda} \left( \begin{array}{cc} 0 & i {c}  \\ -i {c} & 0 \end{array} \right)
\ee 
where the splitting ratio is 
\be{3324}
\lambda \ \ = \ \ \frac{c_1}{c_1+c_2}
\ee
which should be contrasted with 
the stochastic expression Eq.(\ref{e6667}).
If we had ${\lambda=1}$ it would be the same 
problem as single path crossing.
The multiple path geometry is reflected 
in having  ${\mathcal{I} \mapsto \lambda  \mathcal{I}}$
and consequently ${\mathcal{Q} \mapsto \lambda  \mathcal{Q}}$. 
It follows that     
\be{0}
\langle \mathcal{Q} \rangle \ \ &=& \ \  \lambda p
\\ 
\label{e3325}
\mbox{Var}(Q)  \ \ &=& \ \  \lambda^2 {(1-p) p} 
\ee
The latter expression for the variance should 
be contrasted with the stochastic expression Eq.(\ref{e6666}).
We now turn to discuss two surprises 
that are associated with the above results.

The first surprise comes out from Eq.(\ref{e3324}): 
it is the possibility to have $\lambda$ outside of the range ${[0,1]}$. 
This happens if $c_1$ and $c_2$ are opposite in sign 
and close in absolute value.  So we can cook a cycle 
such that the splitting is into $700\%$ in one 
path and $-600\%$ in the other. What does it mean? 
After some reflection one realizes that a proper 
way to describe the dynamics is to say that  
the driving $\dot{u}$ induces a circulating current 
in the system. Using a classical-like phrasing  
one may argue that during the transition the particle 
can encircle the ring 6 times before it makes 
the final crossing to the right side.

The second surprise comes out from Eq.(\ref{e3325}): 
it is the way in which $\lambda$ 
appears in the variance calculation.   
Consider for example a strict adiabatic process 
with ${p=1}$. If say $\lambda=1/2$ we do not get ${\mbox{Var}(Q)=(1/2)^2}$
but rather ${\mbox{Var}(Q)=0}$. One may say that 
we do not have an incoherent partitioning of the 
current, but rather a noiseless exact splitting 
of the wavepacket. This should be contrasted with the 
common picture of getting shot noise due to partition 
of current in open geometries.

\section{Quantum stirring}

We can use the results that have been obtained 
in previous sections in order to calculate $Q$  
for the full stirring cycle which is described in Fig.2. 
For this purpose we regard the full stirring cycle  
as a sequence of two Landau Zener crossings, 
where the first is characterized 
by a splitting ratio $\lambda_{\circlearrowleft}$  
while the second is characterized 
by a different splitting ratio $\lambda_{\circlearrowright}$. 
The net effect is 
\be{4534}
\langle \mathcal{Q} \rangle  
\ \ = \ \  
\lambda_{\circlearrowleft}- \lambda_{\circlearrowright}
\ \ + \ \ \mathcal{O}\left( P_{\tbox{LZ}} \right)
\ee
An optional way to derive this result is to make 
a full 3~level calculation using the Kubo formula \cite{pmc}. 
Here we have bypassed the long derivation 
by making a reduction to a multiple path crossing problem.
The naive expectation is to have ${|\mathcal{Q}|<1}$ if the 
valves are leaky. But by playing with the splitting ratio 
we can get ${|\mathcal{Q}|\gg1}$ per cycle. 
In the language of the Kubo formalism \cite{pmc,pml}
this happens if the pumping cycle encloses 
a degeneracy. A large $\mathcal{Q}$ reflects a huge 
circulating current which is induced by the driving.

We can regard the stirring as an induced persistent current.    
Having figured out what is $\langle\mathcal{Q}\rangle$, the next challenge  
in line is to calculate the variance $\mbox{Var}(Q)$.   
For this purpose we regard the full stirring cycle  
as a sequence of two Landau Zener crossings.
The one period evolution operator can be written as
\be{3423}
U(\mbox{\small cycle}) \ \ = \ \ \Big[ T \ U_{\tbox{LZ}}^{\circlearrowright} \ T \Big] 
\ \eexp{-i\bm{\varphi}} \ \Big[U_{\tbox{LZ}}^{\circlearrowleft}\Big]
\ee
We now explain the ingredients of this expression.
The adiabatic approximation for the $U_{\tbox{LZ}}$ of a single 
Landau-Zener crossing is well known (see e.g. \cite{efrat}):
\be{27}
U_{\tbox{LZ}} \ \ \approx \ \ 
\left(\amatrix{
\sqrt{P_{\tbox{LZ}}} & -\sqrt{1{-}P_{\tbox{LZ}}} \cr   
\sqrt{1{-}P_{\tbox{LZ}}} & \sqrt{P_{\tbox{LZ}}} 
} \right)
\ee
In the strict adiabatic limit ${P_{\tbox{LZ}}=0}$ 
and we denote the respective matrix by $U_{\tbox{LZ}}^{(0)}$.
In the first and second crossings ${P_{\tbox{LZ}}}$ 
might be different and accordingly 
we use the notations $U_{\tbox{LZ}}^{\circlearrowleft}$ 
and $U_{\tbox{LZ}}^{\circlearrowright}$. 
The diagonal matrix $\bm{\varphi}=\mbox{diag}\{\varphi_{+},\varphi_{-}\}$ 
contains the dynamical phases that are accumulated in the 
upper and lower levels during the time between the two crossings, 
and we use the notation $\tilde{\varphi}$ for the phase difference.
The transposition operator~$T$ is required 
because in the second half of the cycle the roles 
of the lower and the upper states are interchanged.
The expression for the total Landau-Zener transition 
probability is an interference of the two possible 
ways to get to the upper level, either by making 
the transition in the first crossing or in the second crossing:
\be{4598}
p \ \ \approx \ \ 
\Big|
\sqrt{P_{\tbox{LZ}}^{\circlearrowleft}} 
- \eexp{i\tilde{\varphi}}
\sqrt{P_{\tbox{LZ}}^{\circlearrowright}}
\Big|^2
\ee
Physically $p$~is the probability {\em not} 
to come back to the initial site. The strict adiabatic 
limit is ${p=0}$. 
 
We turn now to calculate $\mathcal{Q}$ in leading order  
for  ${P_{\tbox{LZ}} \ll 1}$.  
The operator $\mathcal{Q}_{\tbox{LZ}}^{\circlearrowleft}$ is obtained 
by integrating over ${\mathcal{I}(t)=U(t)^{\dagger}\mathcal{I}U(t)}$
with $U(t)=U_{\tbox{LZ}}^{\circlearrowleft}(t)$. 
But in the second half of the cycle $U(t)$ 
is given by Eq.(\ref{e3423}), 
with $U_{\tbox{LZ}}^{\circlearrowright}$ replaced  
by $U_{\tbox{LZ}}^{\circlearrowright}(t)$. 
Consequently we get 
\be{-1}
\int \mathcal{I}(t)dt 
\ \approx \
\mathcal{Q}_{\tbox{LZ}}^{\circlearrowleft}
\ - \
[ T \eexp{-i\bm{\varphi}} U_{\tbox{LZ}}^{\circlearrowleft}]^{\dagger}
\ \mathcal{Q}_{\tbox{LZ}}^{\circlearrowright}
\ [ T \eexp{-i\bm{\varphi}} U_{\tbox{LZ}}^{\circlearrowleft}]
\ee
The first term in this expression is Eq.(\ref{e5555})
multiplied by $\lambda_{\circlearrowleft}$,  
with the $p$ of Eq.(\ref{e543}), and with ${\phi=-\pi/2}$
corresponding to the phase convention in Eq.(\ref{e27}).
Then the second term in this expression becomes   
\be{0}
\lambda_{\circlearrowright}
\left(\amatrix{
-(1-\delta p) & i\delta q \cr   
-i \delta q & +(1-\delta p) 
} \right)
\ee
where 
\be{0}
\delta p 
& \ = \ & 
+2P_{\tbox{LZ}}^{\circlearrowleft}
+P_{\tbox{LZ}}^{\circlearrowright}
- 2\sqrt{P_{\tbox{LZ}}^{\circlearrowleft}P_{\tbox{LZ}}^{\circlearrowright}}\cos(\tilde{\varphi})
\\ \label{e3030}
\delta q 
& \ = \ & 
-2\sqrt{P_{\tbox{LZ}}^{\circlearrowleft}} 
+ \sqrt{P_{\tbox{LZ}}^{\circlearrowright}}\ \exp(i\tilde{\varphi})
\ee
It is easily verified that for $\lambda=1$
we have indeed an agreement with the restricted 
quantum-classical correspondence relation 
of Eq.(\ref{e1515}) where $p$ is given by Eq.(\ref{e4598}). 
For $\lambda\ne1$ the result for the 
expectation value~$\langle \mathcal{Q} \rangle$ 
is in agreement with Eq.(\ref{e4534}).
For the variance we get 
\be{6161}
\mbox{Var}(Q) 
\ \ \approx  \ \ 
\Big|  
\tilde{\lambda}_{\circlearrowleft}  
\sqrt{P_{\tbox{LZ}}^{\circlearrowleft}} 
+
\eexp{i\tilde{\varphi}}
\lambda_{\circlearrowright}
\sqrt{P_{\tbox{LZ}}^{\circlearrowright}} 
\Big|^2 
\ee
where ${\tilde{\lambda}_{\circlearrowleft}=\lambda_{\circlearrowleft}-2\lambda_{\circlearrowright}}$.  
This is a generalization of a result that we had obtained 
using an adiabatic formalism in a previous publication \cite{cnb}. 
The appearance of $\tilde{\lambda}_{\circlearrowleft}$ 
instead of $\lambda_{\circlearrowleft}$ reflects 
the definite site preparation at ${t=0}$,
and therefore it did not emerge in the adiabatic calculation 
where ${t=0}$ has no special meaning.
Notice that the adiabatic approximation 
with ${\tilde{\lambda}_{\circlearrowleft}=\lambda_{\circlearrowleft}}$ 
is formally obtained by replacing $U_{\tbox{LZ}}$ with $U_{\tbox{LZ}}^{(0)}$.

One should realize that the interference expresses 
itself differently in the expressions for~$p$ 
and for $\mbox{Var}(Q)$. 
One may re-phrase this observation by saying 
that for a 3-site ring geometry, 
unlike the case of a 2-site geometry, 
there is no trivial relation between  
the counting statistics and the occupation statistics.

\section{Long time statistics of induced currents}

In this section we would like to consider the long time 
behavior of the counting statistics for either 
non-driven or periodically driven systems.  
In the latter context 
it is convenient to define ${U \equiv U(T)}$ 
as the one-period (Floque) evolution operator, 
and ${\mathcal{Q} \equiv \mathcal{Q}(T)}$  
as the one-period counting operator, where ${T=1}$ is the period. 
The interest is in the counting statistics after~$t$ periods. 
Accordingly Eq.(\ref{Qt}) takes the form  
\be{-1}
\mathcal{Q}(t) 
= \int_0^t \mathcal{I}(t') dt'  
= \mathcal{Q} + U^{-1}\mathcal{Q}U + \cdots +  U^{-t}\mathcal{Q}U^{t}
\ee
It should be clear that in this latter discrete version the 
operator $\mathcal{Q}$ (which is like flow per period), 
plays the same role as the operator $\mathcal{I}$ (flow per unit time). 
For this reason we are not going to duplicate the discussion 
below, and stick to continuous time notations. 
Having defined $\mathcal{I}$ as the current 
trough a specified bond (or more generally 
it would be replaced by the flow per-period), 
we can decompose it into a "DC" part and oscillatory 
part as follows:
\be{0}
\mathcal{I}(t) 
= \sum_{n,m} |n\rangle \eexp{i(E_{n}{-}E_{m})t}I_{nm} \langle m| 
\ \ \equiv \ \ \bar{\mathcal{I}} + \delta \mathcal{I}(t)
\ee
where
\be{0}
\bar{\mathcal{I}}
\ \ \equiv \ \ 
\sum_{E_n{=}E_m} |n\rangle I_{nm} \langle m| 
\ee
In the absence of magnetic fields the stationary 
state of non-driven system is characterized 
by zero "DC" current and we get $\bar{\mathcal{I}}=0$,   
unless the Hamiltonian has a degeneracy. 
For a periodically driven systems $n$ are defined as the eigenstates of 
the Floque operator and in general we 
have $\bar{\mathcal{I}}\ne0$. Accordingly 
\be{-1}
\mathcal{Q}(t) = t\bar{\mathcal{I}}  
+ \sum_{E_n\ne E_m} |n\rangle \left[i\frac{1-\eexp{i(E_{n}{-}E_{m})t}}{E_n{-}E_m}\right] I_{nm} \, \langle m| 
\ee
The non-zero elements of the oscillatory term are all off-diagonal,  
while those of $\bar{\mathcal{I}}$ may be both diagonal and non-diagonal. 
However, without loss of generality we can choose the $n$~basis 
such that $\bar{\mathcal{I}}$ is diagonal.  
If the preparation is a superposition 
of Floque eigenstates, then both  
the average and the dispersion grow linearly.
Accordingly, in general, the asymptotic value 
of the relative dispersion  ${\sqrt{\mbox{Var}(Q)} / \langle \mathcal{Q}\rangle}$  
does not go to zero. In order to make it 
go to zero we have to especially prepare 
the system in an~$n$ eigenstate. 
For such special preparation $\langle \mathcal{Q}(t) \rangle$ 
grows linearly while the dispersion is oscillating 
around a constant value:
\be{-1}
\mbox{Var}(Q) 
& \ \ = \ \ &  
\sum_m \frac{2|I_{nm}|^2}{(E_n-E_m)^2} [1-\cos((E_n{-}E_m)t)] 
\\ 
& \ \ = \ \ & 
\overline{\mbox{Var}(Q)} + \mbox{oscillations}
\label{varQ}
\ee
The time averaged value of the dispersion  $\overline{\mbox{Var}(Q)}$ is given 
above by the first term (without the cosine). The time averaged value 
is not always of physical interest. If only two levels are involved, 
then $\mbox{Var}(Q)$ drops to zero periodically. 
This applies also if several levels are involved, 
as long as their spacing differences are not resolved.  
The example in the following section demonstrates these considerations.

\section{Example for long time counting statistics}

The simplest way to illustrate the discussion of the 
previous section is to consider the counting statistics 
which is associated with the persistent current in a clean ring. 
For simplicity we consider $N$~site ring. The position 
eigenstates are labeled as ${x=0,1,2,...}$, 
while the momentum eigenstates are labeled 
as ${p=k_n=(2\pi/N)n}$ with ${n=0,\pm1,\pm2,...}$.   
The Hamiltonian is 
\be{0}
\mathcal{H} =  -cD-cD^{-1} = -2c\cos(p) 
\ee
where ${D=\exp(-ip)}$ is the one site displacement operator, 
and $c$ is the hopping amplitude. The eigen-energies 
are ${E_n=-2c\cos(2\pi n/N)}$. 
Thanks to the ${n \mapsto -n}$ degeneracy of the 
spectrum the ring can support a persistent current even 
in the absence of a magnetic field.
The velocity operator is a three diagonal matrix
\be{0}
v = i[\mathcal{H},x] = icD-icD^{-1} = 2c\sin(p) 
\ee
We measure the current through the ${0\mapsto1}$ bond. 
Accordingly 
\be{0}
\mathcal{I} \ \ = \ \ -ic\Big[|1\rangle\langle 0| - |0\rangle\langle 1|  \Big]
\label{I_bond}
\ee
We realize that 
\be{0}
[\mathcal{I}(t)]_{nm} 
\ \ = \ \ -ic\frac{1}{N}\Big[\eexp{ik_n}-\eexp{-ik_m}\Big]\,\eexp{i(E_n-E_m)t}
\ee
Upon time averaging only the $n=m$ terms survive and 
they equal ${-i(1/N)2c\sin(k_n)}$. 
Thus we have the identification 
\be{0}
\bar{\mathcal{I}} = \frac{1}{N}v
\ee
If the energies were {\em equally} spaced 
with some level spacing~${\Delta=(2\pi/N)v_{\tbox{F}}}$,  
the motion of the particle would be 
strictly periodic. The period 
of the motion ${2\pi/\Delta}$ would be the time 
to make one round along the ring. 
In such case one easily realizes  
that the variance $\mbox{Var}(Q)$ becomes 
zero at the end of each period.
But if one takes the true dispersion 
into account, one realizes that this periodicity 
is not strict:  after some time 
the quasi-periodic (rather than periodic) 
nature of the motion is exposed. 
The long time average of the fluctuating  
variance can be calculated using Eq.(\ref{varQ}), 
leading to:
\be{0}
\overline{\mbox{Var}(Q)} =
\sum_{m\not=\pm n}^N \frac{2|I_{nm}|^2}{(E_n{-}E_m)^2} 
\approx
\frac{1}{\pi^2} \sum_{\nu=1}^{\infty}\frac{1}{\nu^2} 
= \frac{1}{6} 
\ee
Thus, even if a particle is prepared in a definite  
stationary velocity eigenstate, still the counting 
at the end of a period does not yield a {\em certain} result. 
This uncertainty is related to the non-linearity of the spectrum.

\section{Counting, spreading, and dissipation}

The analysis of the long time behavior of the counting 
statistics along the lines of the previous section is not 
very illuminating once we turn to consider driven systems 
of greater complexity. Technically this is because 
the quasi-energies, unlike~$E_n$ become dense in 
the $[0,2\pi]$ interval. It is therefore more illuminating 
to observe that in the latter case the theory of counting 
is strongly related to the theory of spreading in real 
or in energy space.

Let us assume that we have a ring, 
and that the current is measured through a section~$x$. 
[In the tight binding model we can associate 
a location $x_i$ with each site, and Eq.(\ref{I_bond}) 
is an example for an expression for the current 
through a section at ${x_0 < x < x_1}$]. 
For a ring of length $L$ it is convenient to re-define   
the current operator as:
\be{0}
\mathcal{I}
\ \ := \ \ \frac{1}{L}\int_0^L dx \, \mathcal{I}
\ \ = \ \ \frac{1}{L}v
\ee
which is essentially the velocity operator.
Thus we get a relation between the counting 
operator and the spreading in real space 
\be{0}
\mathcal{Q}(t) \ \ = \ \ \frac{1}{L}(x(t)-x(0))
\ee
where $x(t)$ is the non-periodic extension of the position operator. 
This procedure is nice but it should be clear 
that the re-definition of the current operator 
implies that possibly important information is lost.

More generally we can define an ``associated" 
physical problem as follows. 
Assume that the particles are charged 
(for simplicity we set $e=1$),  
and the current is driven by an electro-motive-force (EMF) 
which is induced by a vector potential 
\be{0}
A(x,t) \ \ = \ \ \Phi(t)\,\delta(x-x_0)
\ee
Then the rate of energy absorption is 
proportional to the current and to the EMF ($-\dot{\Phi}$)
\be{0}
\frac{d\mathcal{H}}{dt} 
\ \ = \ \ \frac{\partial \mathcal{H}}{\partial t} 
\ \ = \ \ \frac{\partial \mathcal{H}}{\partial \Phi} \dot{\Phi} 
\ \ = \ \ - \dot{\Phi} \times \mathcal{I} 
\ee
Thus the transported charge $\mathcal{Q}$ implies energy absorption   
\be{0}
\mathcal{Q}_{\tbox{absorption}} 
\ \ \equiv \ \ 
(\mathcal{H}(t){-}\mathcal{H}(0))
\ \ = \ \ 
-\dot{\Phi} \times \mathcal{Q} 
\ee
The energy absorption can be either positive or negative, 
but if the system is chaotic it can be argued 
that its dispersion, and hence the average are growing 
as a function of time.  The same applies to $\mathcal{Q}$. Namely:
\be{0}
\mbox{Var}(Q) \ \ = \ \ \int_0^t\int_0^t  \langle \mathcal{I}(t)\mathcal{I}(t') \rangle dtdt'
\ee
Thus both the counting statistics and the dissipation 
reflect the fluctuations of the current (Kubo). 
This point of view is quite powerful. Instead of 
thinking about ``counting statistics" (which is quite 
abstract) we can think about ``energy absorption" 
for which we have better intuition and better 
theoretical tools for analysis.

\section{Summary}

In closed geometries counting statistics does not obey the common 
stochastic point of view, but rather reflects the coherent nature 
of the quantum transport. For a single path crossing in a 2~site 
system the first and second moments coincide with the naive classical expectation 
due to a restricted quantum-classical correspondence principle 
that can be established. But more generally, in multiple path 
geometries, the results are not as naively expected. 
 
In a double path geometry the particle has two paths 
of transport. The splitting ratio $\lambda$ would be 
in the range $[0,1]$ if the particle were classical. 
But in the case of coherent transport $\lambda$ can be 
any number (either negative or positive). 
We have explained what is the correct way 
to incorporate the splitting ratio $\lambda$
in the calculation of the counting statistics.

During a double path crossing the probability amplitude of the particle 
is transported {\em simultaneously} via the two available paths.
The coherent splitting is ``exact", 
and consequently~$Q$ has zero variance. 
If the particle had finite probabilities to go {\em either} 
via the first path {\em or} via the second path, 
it would imply a non-zero variance (which is not the case). 

Possibly the most interesting result is the analysis 
of a full stirring cycle where the counting statistics 
becomes unrelated to the changes in the occupation 
statistics. In particular we showed that the interference 
of sequential Landau-Zener crossings is reflected 
differently in the respective expressions 
for  ${\mbox{Var}{Q}}$ and for~$p$.

The RMS of the fluctuations of $\mathcal{Q}$ 
in a coherent stirring process 
grows like ${\propto t}$ and not like ${\propto \sqrt{t}}$. 
This linear increase can be avoided, or at least 
minimized, if there is control over the preparation 
of the system. Then we are left with a constant residual dispersion. 
For a driven system the counting statistics is related 
to the study of spreading and dissipation in energy space,  
and hence the growth of the variance constitutes 
a reflection of linear response characteristics. 


\section*{Acknowledgments}
 
The research was supported by a grant from the DIP, 
the Deutsch-Israelische Projektkooperation, 
and by a grant from the USA-Israel Binational Science Foundation (BSF).


\end{document}